\documentclass{aa}  
\usepackage{lineno}
\usepackage{pifont}%
\newcommand{\cmark}{\ding{51}}%
\newcommand{\xmark}{\ding{55}}
\usepackage{graphicx}
\usepackage{amsmath}
\usepackage{txfonts}
\usepackage{lipsum}
\usepackage{subcaption}      
\usepackage{comment}
\usepackage{xcolor}
\usepackage{makecell}
\definecolor{cornflowerblue}{RGB}{100, 149, 237} 
\definecolor{orange}{RGB}{255,127,14}
\definecolor{darkorchid}{RGB}{153,50,204}
\usepackage{dashrule}
\usepackage[colorlinks=true, linkcolor=blue, citecolor=blue, urlcolor=black]{hyperref}
\usepackage[font=small,labelfont=bf]{caption}
\usepackage{multicol}
\usepackage{multirow}
\usepackage[normalem]{ulem}
\usepackage{enumitem}

\usepackage{lscape}             
\usepackage{placeins}           
                                
\begin{document}

   \title{Cored galaxies in cuspy dark matter halos}


   \author{Fernando Valenciano\inst{1,2}\thanks{ \texttt{fernando.valenciano@iac.es}}
        \and Jorge Martin Camalich\inst{1,2} 
        \and Arianna Di Cintio\inst{2,1}
        \and Julio F. Navarro\inst{3} 
        \and Giuseppina Battaglia\inst{1,2}
        \and Rapha\"el Errani\inst{4}
        \and Justin I. Read\inst{5}}

   \institute{Instituto de Astrof\'{\i}sica de Canarias, C/ V\'{\i}a L\'{a}ctea, s/n, E-38205 La Laguna, Tenerife, Spain 
         \and
Departamento de Astrof\'{\i}sica, Universidad de La Laguna, Avenida Francisco S\'{a}nchez, s/n, E-38205 La Laguna, Tenerife, Spain
\and Department of Physics and Astronomy, University of Victoria, Victoria, BC, V8P 5C2, Canada
\and McWilliams Center for Cosmology and Astrophysics, Dept. of Physics, Carnegie Mellon University, Pittsburgh, PA 15213, USA
\and University of Surrey, Physics Department, Guildford, GU2 7XH, UK}

 
\abstract
{}      
{We investigate constraints on the inner stellar density profile from photometric data of dwarf spheroidal and ultra-faint dwarf galaxies.
Our aim is to clarify under what conditions cored stellar profiles require dark matter halos that are also cored, deviating from the cuspy profiles expected for cold dark matter halos.}
{We consider a variety of spherically symmetric stellar profiles, which we classify  as ``strong'' or ``weak'' cores and cusps according to the behavior of the slope ($b_0$) and logarithmic slope ($\gamma_0$) at their centers. We  explore which profiles lead to unphysical negative distribution functions when embedded in a cuspy halo, treating isotropic and anisotropic kinematics separately.}
{We find that weakly-cored stellar profiles in 3D (i.e., $b_0 \neq 0$, $\gamma_0=0$) can be consistent with cuspy dark matter profiles, but strong 3D cores ($b_0=\gamma_0=0$) are not. However, both weak and strong 3D cores yield nearly indistinguishable inner profiles in projection, which implies that ruling out a dark matter cusp from photometric data alone is highly challenging. As an example, we study the profiles of  ultra-faint dwarf galaxies and find that they are consistent with both weak and strong 3D cores. This is not just a result of the limited numbers of stars in these systems, since we reach the same conclusion even for Fornax, one of the most luminous and best-studied dwarf spheroidal companions of the Milky Way.
}
{We conclude that, based on current data and analysis techniques, cored surface density profiles in nearby dwarf galaxies cannot be taken as strong evidence against the presence of cuspy dark matter halos.}

\maketitle
\nolinenumbers

\section{Introduction}

The hierarchical structure formation of dark matter (DM) halos is a successful framework to describe the history of accretion and merger events, giving rise to a complex clustering evolution of halos and subhalos. 
Within this paradigm, the Navarro-Frenk-White (NFW) density profile~\citep{1996ApJ...462..563N,1997ApJ...490..493N} remains the canonical form to describe DM halo properties derived from collisionless cold dark matter (CDM) $N$-body simulations, featuring a universal inner slope $\rho \propto r^{-1}$. However, stellar and gas kinematics in dwarf galaxies often favor flat density profiles at the center instead of ``cuspy'' ones~\citep{1994ApJ...427L...1F,1994Natur.370..629M,2002A&A...385..816D,2011AJ....141..193O,Iorio_2016,2017MNRAS.467.2019R}. 
This discrepancy is known as the ``core-cusp problem'', and is one of a number of so-called small-scale challenges to $\Lambda$CDM~\citep{2017ARA&A..55..343B}.  

The proposed solutions range from modifications of the fundamental nature of DM, such as fuzzy DM~\citep{Hu:2000ke,Schive:2014dra,Hui:2016ltb}, warm DM~\citep{Sommer-Larsen:1999otf,Bode:2000gq}, and self-interacting DM~\citep{Spergel:1999mh,Tulin:2013teo,Tulin:2017ara}, to baryonic processes within $\Lambda$CDM, including the flattening of cusps into cores through supernova feedback~\citep{Navarro:1996bv,Pontzen:2011ty,DiCintio:2013qxa,Tollet:2015gqa,Read:2015sta}.  
To distinguish between proposed solutions, it is widely recognized that one must focus on the smallest galaxies, near the threshold of galaxy formation ($\sim 10^5-10^6 M_{\odot}$)~\citep{DiCintio:2013qxa,2015MNRAS.453.2133B,Read:2015sta}. 
In this regime, CDM and more exotic scenarios predict markedly different profiles; however, gathering the number of kinematic tracers needed for precise inference of $\rho(r)$ in this regime is extremely challenging.

Less attention has been devoted to the analysis of the spatial distributions of the tracers themselves. In particular, it remains unclear whether the form of the potential determines the distribution of stars and how this information may be used to infer its properties. A standard framework to approach this problem is the analysis of stationary solutions of the collisionless Boltzmann equation, described by the tracer distribution functions (DF).

It is well known that, via the Eddington inversion formalism~\citep{1916MNRAS..76..572E,2008gady.book.....B}, not all combinations of gravitational potentials and tracer densities correspond to physical (positive) DFs. These arguments of mathematical consistency have been used for constructing mock galaxies~\citep[e.g.][]{2004ApJ...601...37K,2006ApJ...642..752A,2020MNRAS.491.4591E,2025arXiv250219475E,2024ApJ...968...89E} and have been pushed further to suggest that cored stellar surface-density profiles are inconsistent with embeddings in cuspy DM halos~\citep{2023ApJ...954..153S,2024A&A...690A.151S,2024RNAAS...8..167S,2025A&A...694A.283S}, see also ~\citet{2006ApJ...642..752A,2013A&A...558L...3B,2018JCAP...09..040L}. More recently, the same arguments have been invoked in the context of ultra-faint dwarf galaxy (UFD) observations to question the validity of $\Lambda$CDM~\citep{2024ApJ...973L..15S}.

These inferences are drawn from surface density profiles, whereas the DF consistency constraints are formulated in terms of the compatibility between the volumetric densities of DM and stars. Previous studies have shown that minor differences in the stellar surface profiles (2D) may correspond to substantially different stellar 3D densities, energy distributions and DFs~\citep{2024ApJ...968...89E}. It is therefore crucial to start by scrutinizing the relation between three-dimensional (3D) densities and their projected two-dimensional (2D) counterparts, and subsequently assessing how DF-based consistency conditions can be applied to observations that are limited to the latter.

In Sec.~\ref{sec:formalism}, we introduce a classification of spherically symmetric 3D density profiles based on their properties, and those of their 2D projections, at the center. We also present the DF formalism, with the Eddington equation as its cornerstone, and explore the connections between our classification of profiles and the mathematical consistency of their embeddings within cuspy DM halos. In Sec.~\ref{sec:apps}, we apply the formalism to different density profiles and DFs and use these models to fit observational data. In Sec.~\ref{sec:conclusions}, we discuss the main conclusions of this work.

\section{Mathematical formalism}
\label{sec:formalism}
\subsection{Types of cores and cusps}
\label{sec:cores&cusps}
We introduce a classification scheme designed to capture the variety of 3D density profiles $\rho(r)$ that can be constructed under the assumptions of spherical symmetry and that decrease monotonically with radius $r$. One defines the slope and log-slope of the distribution,
\begin{align}
\label{eq:slope} 
    b(r)&=-\frac{d\rho}{dr} &\text{ with  
  }\;\;\;\; b_0=\lim_{r\to0} b(r),
\end{align}
\begin{align}
\label{eq:logslope} 
\gamma(r)&= -\frac{d\log\rho}{d\log r} &\text{ with  
}\;\;\;\;\; 
\gamma_0=\lim_{r\to0} \gamma(r),
\end{align}
where $\gamma(r)=r\,b(r)/\rho(r)$. Let us focus on the subset of volumetric densities $\rho(r)$ that are continuous at $r=0$; i.e. finite at the center. We assume that these distributions can be approximated at $r\to0$ as
\begin{align}
\label{eq:expansion1}
\rho(r)&\approx\rho_0\left(1-a r^\alpha\right)&\;\;\;\;\text{ with }\;\;\;\;\alpha>0\\
b(r)&\approx r^{\alpha-1},
&\gamma(r)\approx r^\alpha.
\end{align}
with $\rho_0$ and $a$ positive dimensionful numbers. The log-slope at the center is $\gamma_0=0$ for all $\alpha>0$, while the value of the slope depends on $\alpha$. We \textit{define} these distributions as ``cores'' and introduce the following subclassification:
\begin{itemize}
\item\textbf{Strong core:} density profile finite at $r\to0$ with $\alpha>1$ and $b_0=\gamma_0=0$.
\item \textbf{Weak core:} density profile finite at $r\to0$ with $0<\alpha\leq1$ and $b_0\neq0$, $\gamma_0=0$.
\end{itemize}
Note that all these profiles are cored according to the log-slope $\gamma$, but the slope $b$ distinguishes between truly flat cores and shallow ones.

In contrast, we define as ``cusps'' volumetric densities that are divergent at $r\to0$. For example, they can diverge in a power-law form,
\begin{align}
\label{eq:expansion2}
\rho(r)&\approx\frac{\rho_s}{r^{\gamma}}&\text{ with }\;\;\;\;\gamma>0,\\
b(r)&\approx\frac{1}{r^{\gamma+1}},
&\gamma(r)\approx\gamma
\end{align}
with $\rho_s$ a positive dimensionful number. The log-slope around $r\approx0$ is the exponent $\gamma\neq0$ while the slope is singular at $r\to0$. Distributions can also become logarithmically divergent at the center,
\begin{align}
\label{eq:expansion3}
\rho(r)&\approx\rho_s\log\frac{1}{r},&\\
b(r)&\approx\frac{1}{r},
&\gamma(r)\approx0.
\end{align}
We extend our classification to singular density profiles:
\begin{itemize}
\item \textbf{Weak cusp:} density profile divergent at $r\to0$ with power $0\leq\gamma<1$. 
\item \textbf{Strong cusp:} density profile divergent at $r\to0$ with power $\gamma\geq1$. 
\end{itemize}
The relevance of this classification will become clear when we discuss projection effects and the consistency conditions for the corresponding DFs.

\subsection{Surface density profiles}
\label{sec:SBprofiles}

The 2D surface density profile corresponding to $\rho(r)$ is defined by the Abel projection,
\begin{equation}
\label{eq:projSigma}
    \Sigma(R)=2\int^{\infty}_Rdr\frac{\rho(r)r}{\sqrt{r^2-R^2}},
\end{equation}
as a function of the projected radius $R$. The slope and log-slope of $\Sigma(R)$ can also be defined analogously to Eqs.~\eqref{eq:slope} and \eqref{eq:logslope},
\begin{align}
\label{eq:proj_slope}
    \bar{b}(R)&=-\frac{d\Sigma}{dR}&\text{ with  
  }\;\;\;\; \bar{b}_0=\lim_{R\to0} \bar{b}(R),
\end{align}
\begin{align}
\label{eq:proj_logslope}
    \bar{\gamma}(R)&= -\frac{d\log\Sigma}{d\log R} &\text{ with  
  }\;\;\;\; \bar\gamma_0=\lim_{R\to0} \bar\gamma(R).
\end{align} 
We can relate the 3D and 2D slopes by differentiating Eq. \eqref{eq:projSigma}
\begin{equation}
\label{eq:slopes_rel}
\bar b(R)=2R\int^\infty_0 dl \frac{b\left(\sqrt{R^2+l^2}\right)}{\sqrt{R^2+l^2}},
\end{equation}
where $l$ is the line-of-sight coordinate and $\bar b$ is determined by the distribution $\rho(r)$ around $r\approx0$~\footnote{Assuming the volumetric density falls fast enough at $r\to\infty$ or is cut off at some radius so that the total mass is finite.} (See Appendix \ref{app:math_vitamins}). 

For strong cores ($\alpha>1$) the integral converges and $\bar b_0=\bar\gamma_0=0$. For weak cores, the behavior around $R\approx0$ is,
\begin{align}
\label{eq:bRs_cores}
\bar b(R)&\approx R^{\alpha}\,,&0<\alpha<1,\nonumber\\
\bar b(R)&\approx R\log\frac{1}{R}\,,&\alpha=1,
\end{align}
and the slopes of the projection are also $\bar{b}_0=\bar{\gamma}_0=0$. Therefore, by extending our classification from volumetric to surface density profiles, we derive the following corollary:\\

\noindent\emph{All spherically symmetric 3D density profiles that are continuous functions at their center (i.e. have finite central density) are strong cores in 2D.}\\

\begin{figure*}[htb!]
    \centering
    \includegraphics[width=1\textwidth]{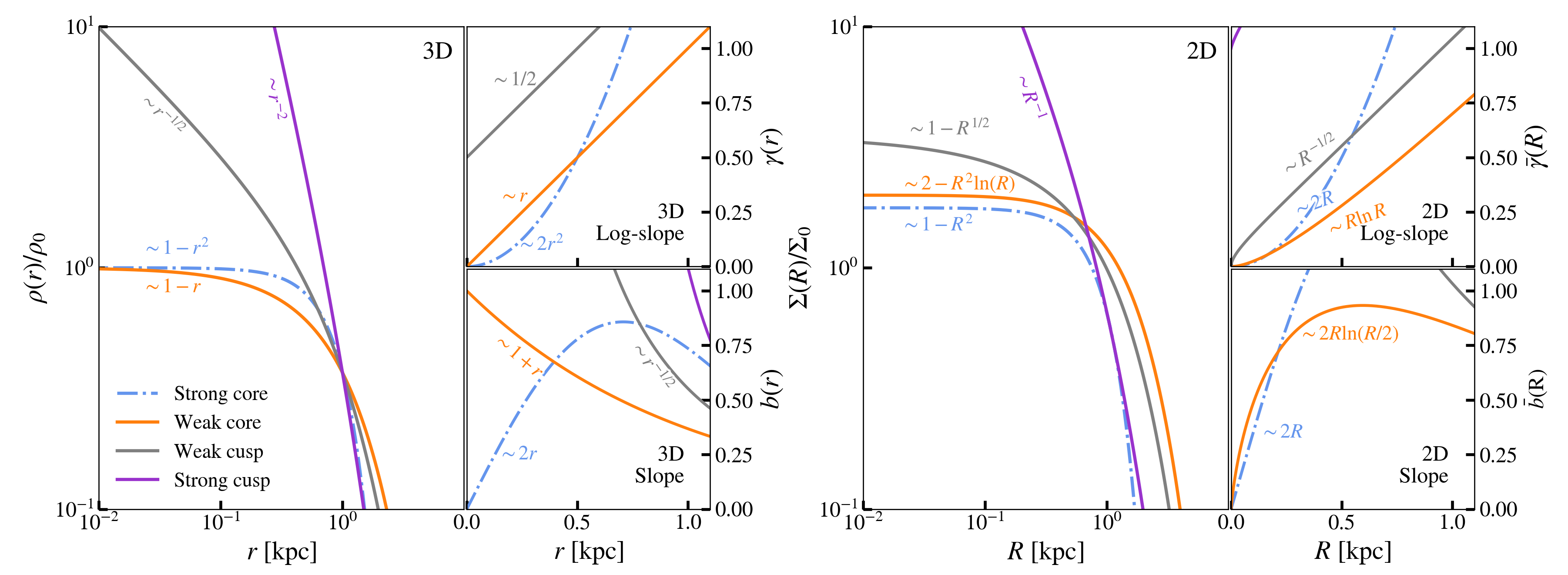}
    \caption{Schematic overview of the proposed classification: strong cores 
    (\textcolor{cornflowerblue}{\rule[0.5ex]{0.2em}{1pt}\hspace{0.2em}\rule[0.5ex]{0.2em}{1pt}\hspace{0.2em}\rule[0.5ex]{0.2em}{1pt}}), weak cores 
    (\textcolor{orange}{\rule[0.5ex]{1em}{1pt}}), weak cusps 
    (\textcolor{gray}{\rule[0.5ex]{1em}{1pt}}) and strong cusps 
    (\textcolor{darkorchid}{\rule[0.5ex]{1em}{1pt}}) labeled by their asymptotic behavior at the center. The left panels show the 
    three-dimensional (volumetric) density profiles, together with the corresponding 
    log-slope $\gamma$ (upper right) and parameter $b$ (lower right). The 
    right panels present the analogous projected (surface–density) quantities. Within 
    this framework, strong and weak cores both appear as strong cores in projection, 
    weak cusps resemble weak cores, and strong cusps remain clearly cuspy.}
    
    \label{fig:classification}
\end{figure*}

\begin{table}
\caption*{\textbf{DF in cuspy DM halos}}
\vspace{-2mm}
    \centering
    \small
  \renewcommand{\arraystretch}{1.7}
\setlength{\arrayrulewidth}{.35mm}
  \setlength{\tabcolsep}{0.25 em}  
    \begin{tabular}{|cc|cc|c|}
    \hline
  \multicolumn{2}{|c|}{$\mathbf{\Sigma}$ (2D - observed)}  &\multicolumn{2}{c|}{$\mathbf{\rho}$ (3D - derived) } & \textbf{$f(E)>0$}  \\
\hline
 \multirow{2}{5em}{\textbf{Strong core}} & \multirow{2}{5em}{$\bar b_0=\bar\gamma_0=0$} &   \textbf{Strong core} &$b_0=\gamma_0=0$&  {\color{red} \xmark}\\
\Xcline{3-5}{0.5pt}
   &  & \textbf{Weak core} &  $b_0\neq0$, $\gamma_0=0$   & {\color{green}\cmark} \\
\Xcline{1-5}{0.5pt}
 \textbf{Weak core} &$\bar b_0\neq0$, $\bar\gamma_0=0$  &\textbf{Weak cusp} &  $0\leq\gamma<1$  &{\color{green}\cmark}   \\
\Xcline{1-5}{0.5pt}
 \textbf{Cusp}   & $\bar\gamma_0=\gamma-1$ & \textbf{Strong cusp} & $\gamma\geq1$ &  {\color{green}\cmark}\\
\hline      
    \end{tabular}
    \vspace{3mm}
    \caption{Classification of surface (observed) and volumetric (derived) density profiles of stars in terms of their properties around the center. The last column indicates whether $f(E)\geq0$ is expected in cuspy DM halos based solely on the condition $d\rho/d\Psi\neq0$.
    }
    \label{tab:class}
\end{table}

In case of 3D profiles that diverge at $r\to0$, the properties around $R\approx0$ depend on the value of $\gamma$,
\begin{align}
\label{eq:sigmas_sing0}
\Sigma(R)&\approx\Sigma(0)-\pi\rho_s R, & \gamma=0,\\
\label{eq:sigmas_sing01}
\Sigma(R)&\approx\Sigma(0)+C\,R^{1-\gamma}, & 0<\gamma<1,\\
\label{eq:sigmas_sing1}
\Sigma(R)&\approx\log\frac{1}{R}, & \gamma=1,\\
\Sigma(R)&\approx\frac{1}{R^{\gamma-1}}, & \gamma>1,
\end{align}
where for $0\leq\gamma<1$ the surface density at the center, $\Sigma(0)$, is a finite number and $C$ is a $\gamma$-dependent number. The slope of the projection of the cusps at $R\approx0$ is
\begin{align}
\label{eq:projslop_cusp}
\bar b(R)&\approx\pi\rho_s,&\gamma=0,\\
\bar b(R)&\approx\frac{1}{R^{\gamma}},&\text{for all }\;\gamma>0,
\end{align}
and their log-slopes are
\begin{align}
\bar \gamma(R)&\approx0,&0\leq\gamma\le1,\\
\bar \gamma(R)&\approx\gamma,&\gamma>1.
\end{align}
Weak cusps in 3D have $\bar b_0\neq0$ and $\bar\gamma_0=0$ and become weak cores in 2D, while strong cusps in 3D remain cuspy in 2D. Thus, one can draw a complementary corollary, which is essentially the inverse of that derived above for the projection of cores:\\

\noindent\textit{2D surface density profiles that are not strong cores necessarily correspond to cuspy 3D volumetric density profiles (assuming spherical symmetry).}\\

Table~\ref{tab:class} and Figure~\ref{fig:classification} provide a schematic overview of the proposed classification, together with the corresponding logarithmic slopes~$\gamma$ and slopes~$b$ in both the 2D and 3D cases.

\subsection{Distribution functions and the Eddington equation}
\label{sec:Eddington}

For spherically symmetric systems of stars with isotropic velocity distribution, the phase-space DF only depends on the binding energy $E$. The volumetric density is:
\begin{equation}  \rho(r)=4\pi\sqrt{2}\int^{\Psi(r)}_0f(E)\sqrt{\Psi(r)-E}\,dE
\end{equation}
The function $\Psi(r)=-\Phi(r)$ encodes the gravitational potential ($\Phi$) of the static DM halo. From this definition,
\begin{equation}
\label{eq:suff_condition}
    \frac{d\rho}{d\Psi}=2\pi\sqrt{2}\int^{\Psi}_0\frac{f(E)}{\sqrt{\Psi-E}}dE,
\end{equation}
which can be inverted to obtain the DF, assuming that the stellar density profile vanishes sufficiently fast at large radii so that $d\rho/d\Psi = 0$ for $\Psi=0$,
\begin{equation}
\label{eq:Eddington}
    f(E)=\frac{1}{\sqrt{8}\pi^2}\int^{E}_0\frac{d^2\rho}{d^2\Psi}\frac{d\Psi}{\sqrt{E-\Psi}}.
\end{equation}
This equation (called Eddington formula) does not guarantee that the derived DF is positive and, thus, physical~\citep{2008gady.book.....B}. 
Within this framework, Eq.~\eqref{eq:suff_condition} provides a sufficient condition for the consistency between a volumetric density $\rho(r)$ and the DM halo potential $\Psi(r)$~\citep{2018JCAP...09..040L}. If $d\rho/d\Psi = 0$, then the DF $f(E)$ cannot remain positive everywhere and the DF must take negative values at some point within the interval $0 \leq E \leq \Psi$. 

This condition can be related to the case $d\rho/dr=0$ when the embedding DM potentials are not strong cores~\citep{2023ApJ...954..153S}.

It is important to emphasize that this sufficient condition for a negative DF applies to 3D densities, not to the projected 2D surface densities accessible to photometric observations. Since weak cores in 3D project to strong cores in 2D, we conclude:\\

\noindent\emph{The existence of a strong core in photometric surface densities is not a sufficient condition to exclude cuspy DM halos.}\\

In any case, we must stress that this is a sufficient and not a necessary condition for $f(E)<0$, i.e. the DF can still be unphysical even if $d\rho/d\Psi \neq 0$. Another sufficient condition for the negativity of the DFs that can be derived directly from Eq.~\eqref{eq:Eddington} is related to the condition $d^2\rho/d\Psi^2 < 0$~\citep{2018JCAP...09..040L}. This can occur for different combinations of stellar profiles and DM embeddings as for example for stellar weak cores that feature a sharp transition between the inner and outer log-slopes. We illustrate this with an example in the Appendix~\ref{sec:d2rho}.

Moreover, the extension of the analysis to anisotropic profiles introduces new aspects in the consistency in the dynamical modeling of dwarf galaxies which are somewhat orthogonal to the ones considered in this paper as they now depend on the radial dependence of the anisotropy ~\cite[see e.g.][]{2006ApJ...642..752A,2023ApJ...954..153S}. We address this using an Osipkov-Merritt anisotropy model~\citep{1979SvAL....5...42O,1985AJ.....90.1027M} in Appendix~\ref{sec:anisotropy}   

\section{Applications}
\label{sec:apps}
In this section we examine a set of stellar-density profile models to assess the proposed classification and to evaluate their compatibility with being embedded within a cuspy DM halo.

\subsection{Profile models}

For the DM halo, we choose a Hernquist profile \citep{1990ApJ...356..359H} 
\begin{equation}
\label{eq:Hernquist_prof}
    \rho_{\rm DM}(r)=\frac{M_h}{2\pi r_s^3}\frac{1}{(r/r_s)(1+r/r_s)^3}
\end{equation}
which offers a simple analytical form for the potential,
\begin{equation}
\label{eq:Hernq_potential}
    \Phi_{\rm DM}(r)=-\frac{G M_h}{r_s}\frac{1}{1+r/r_s}
\end{equation}
where $M_h$ and $r_s$ are the halo mass and scale radius. This DM density profile is a strong cusp (with $\gamma_0=1$), consistent with those of the DM halos in DM-only cosmological simulations~\citep{1996ApJ...462..563N,1997ApJ...490..493N}. Our conclusions do not depend on the form of the embedding potential at large $r$ and  remain unchanged when using a NFW potential (see Appendix \ref{app:NFW}).

For the stellar distributions we consider various volumetric profiles. Combined with the potential of the host DM halo in Eq.~\eqref{eq:Hernq_potential} we infer the DF using Eq.~\eqref{eq:Eddington}.
\vspace{1mm}

\textit{1. Plummer:} We start with the Plummer profile \citep{1911MNRAS..71..460P}, which is an often-used profile with a strong core in 3D,
\begin{equation}
\label{eq:Plummer3D}
    \rho_\star(r) = \frac{3M}{4\pi r_{P}^3}\left(1+\frac{r^2}{r_{P}^2}\right)^{-5/2},
\end{equation}
where $M$ is the total stellar mass and $r_{P}$ is the scale radius that sets the core size. The projected 2D distribution is:
\begin{equation}
\label{eq:Plummer2D}
\Sigma(R)=\frac{M}{\pi r_P^2}\left(1+\frac{R^2}{r_P^2}\right)^{-2}.
\end{equation}

\begin{figure*}[htb!]
    \centering
    \includegraphics[width=0.8\textwidth]{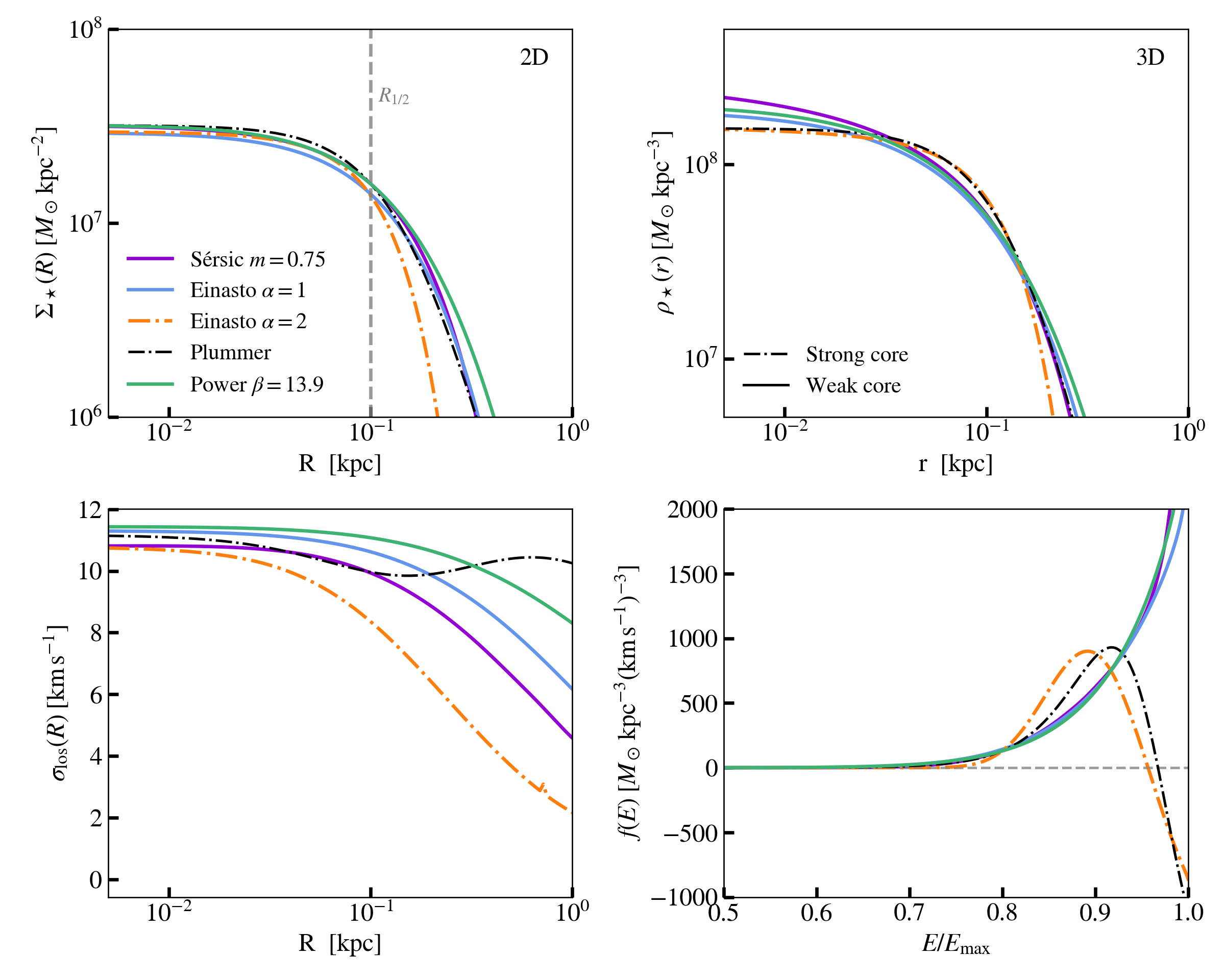}
    \caption{Stellar surface density $\Sigma_\star$, 3D density $\rho_\star$, line-of-sight velocity dispersion $\sigma_\mathrm{los}$ and DF, obtained for various stellar profiles embedded into a DM halo based on the Hernquist potential with total mass $M_{h}=10^9M_{\odot}$ and scale radius $r_s=1$kpc. For the stellar tracers we use $\Sigma_0=3\times10^7M_{\odot}\text{kpc}^{-2}$, $R_{1/2}=0.1$kpc and assume an isotropic velocity distribution. Curves plotted with solid lines (\textcolor{black}{\rule[0.5ex]{1em}{1pt}}) or dot-dashed lines (\textcolor{black}{%
\rule[0.5ex]{0.4em}{1pt}\hspace{0.15em}%
\raisebox{0.5ex}{.}\hspace{0.15em}%
\rule[0.5ex]{0.4em}{1pt}}) correspond to weak or strong core profiles in 3D, respectively.
    }
    \label{fig:models}
\end{figure*}

\textit{2. Einasto:} We also model the stellar distribution with the Einasto profile~\citep{1965TrAlm...5...87E},
\begin{equation}
\label{eq:rho_Einasto}
    \rho_\star(r)=\rho_0\exp\left[-\left(\frac{r}{r_E}\right)^{\alpha}\right],
\end{equation}
with $\alpha\geq 0$. This profile is always finite at $r\to 0$ and illustrates the classification of cores introduced in Sect.~\ref{sec:formalism}, into weak ($0<\alpha\leq1$) and strong ($\alpha>1$) cores. 
The projection of the Einasto profile in 2D has an analytical representation~\citep{2012A&A...540A..70R}, where one can explicitly check that surface density profiles based on the projection of the Einasto profiles are strong cores in 2D. 
\vspace{1mm}

\textit{3. S\'ersic:} A model to describe directly the surface brightness distribution of galaxies is the S\'ersic profile~\citep{1963BAAA....6...41S},
\begin{equation}
\label{eq:sigma_Sersic}
    \Sigma(R)=\Sigma_0\exp\left[-\left(\frac{R}{R_\star}\right)^{\frac{1}{m}}\right].
\end{equation}
The 3D volumetric density profile obtained through the inverse Abel transformation adopts an analytical representation in terms of special functions, and its class depends on the S\'ersic index $m$~\citep{2011A&A...534A..69B}. For $m\geq1$ the 3D distribution is a cusp and for $1/2<m<1$ it is a weak core. The case $m = 1$ corresponds to an exponential surface brightness profile, whose deprojection yields 
$\rho(r) \approx (\Sigma_0 / R_\star)\,\ln(1/r)$ near $r \approx 0$; that is, a weak cusp with a central log-slope $\gamma_0=0$. 
At the other end of the weak-core regime, the limiting case $m = 1/2$ deprojects to a Gaussian density distribution in 3D that is a strong core. Finally, for $m\leq1/2$ the S\'ersic profile also yields a volumetric distribution that is a strong core but that is not a monotonically decreasing function of the radius~\citep{2001MNRAS.321..269T}, expected to lead to a negative DF because they can verify $d^2\rho/d\Psi^2$<0.
\vspace{1mm}

\textit{4. Power-potential:} One can also build a stellar profile using a top-down approach, by starting from a definite positive DF. The simplest example is to define a mono-energetic DF \citep{2024ApJ...968...89E,2025ApJ...992..162E}, i.e.
\begin{equation}
\label{eq:DFmonoenergetic}
f(E)\propto\,\delta(E-E_0),
\end{equation}
where $E_0$ denotes the binding energy of the stellar component within the DM halo. For this simple form of DF, the resulting stellar density profile is readily obtained from the underlying DM potential~\citep[Eq.~(13) in][]{2025ApJ...992..162E}
\begin{equation}
    \rho_\star(r) = \left\{
    \begin{array}{lr}\rho_0\sqrt{(\Psi(r) - E_0)/(\Psi(0)-E_0) }&\text{if $\Psi(r) \geq E_0 $}, \\
    0 &\text{if $\Psi(r) < E_0,$} \end{array}
    \right.
\end{equation}
where we are assuming that the potential is finite at $r=0$. Moreover, this potential is also related to the DM density distribution via the Poisson equation. Assuming that the gravitational potential behaves at the center as
\begin{align}
&\Psi(r)\approx \Psi(0)\left(1-a_\Psi  r^{2-\gamma}\right), & \text{with}\;\;\;0\leq\gamma<2, \label{eq:gravpot_center}  
\end{align}
one obtains that, for $r \approx 0$, 
\begin{align}
 & \rho_\star(r)\approx\rho_0(1-a_\star r^{2-\gamma}),&\rho_{\rm DM}(r)\approx r^{-\gamma}. \label{eq:rhostarvsrhoDM}
\end{align}
Since any DF may be expressed as the continuous limit of a sum of mono-energetic DFs, this can provide a direct link between the DM density profile and the stellar density profiles that can be sustained in dynamical equilibrium. 

In particular, cuspy DM halos with $\gamma\ge 1$ can sustain stellar weak cores but not strong cores. 
In the case of a Hernquist DM distribution, the mono-energetic stellar DF corresponds to
\begin{align}
\rho_\star(r)=\left\{
    \begin{array}{lr}
       \rho_0\,\left(\frac{r_0}{r_s}-1\right)^{-1/2}\left(\frac{r_0}{r+r_s}-1\right)^{1/2} & \text{if } r\leq r_0-r_s,\\
        0\;\;\; & \text{if } r>r_0-r_s,
    \end{array}\right.\;\;\;,
\end{align}
where $r_0=GM_h/E_0$. The profile is finite at its center, with asymptotic behavior around $r\approx0$,
\begin{align}
\rho_\star(r)\approx\rho_0\left(1-\frac{r_0}{2r_s(r_0-r_s)}\,r\right), 
\end{align}
explicitly showing that it is a weak core in 3D. Finally, to sustain a stellar strong core with $\gamma_0=0$, from Eq.~\eqref{eq:rhostarvsrhoDM} one concludes that the DM mass distribution must be cored~\footnote{For example, stellar strong cores in 3D are obtained from the mono-energetic DF if the potential corresponds to a Plummer distribution of DM, $\Phi(r)=-GM_h/\sqrt{r^2+r_P^2}$.}.

Another ansatz for the DF is the function  
\begin{align}   
\label{eq:DF_BU}
f(E)&=F E^{\,\beta-3/2}, & E >0,
\end{align}
which can be integrated to obtain the density profile  analytically~\citep{2008gady.book.....B},
\begin{equation}
\label{eq:dens_BU}
    \rho_\star(r)=c_\beta\Psi(r)^{\beta}, \hspace{4mm} (\Psi>0),
\end{equation}
hereafter the ``power-potential'' distribution. 
The $c_\beta$ is a constant proportional to $F$ in Eq.~\eqref{eq:DF_BU} that in order to be finite requires $\beta>\frac{1}{2}$. In case the potential is a Hernquist model, this DF leads to 
\begin{equation}
    \rho_\star(r)=\frac{\rho_0}{(1+r/r_s)^\beta},
\end{equation}
where $\rho_0=c_\beta (GM_h/r_s)^{\,\beta}$. A global factor $\propto F$ determines the central density, while $\beta$ determines the radius scale of the star distribution relative to that of the potential $r_s$.~\footnote{Note that this is an incarnation of Zhao's general volumetric density \citep{1996MNRAS.278..488Z} with power-law parameters $(1,\beta,0)$.} 

The top-down approach, illustrated by the mono-energetic and power-potential profiles, provides a simple way to construct stellar distributions embedded in cuspy DM halos that are strong cores in 2D and with a DF that, by construction, is positive in all the binding energy domain. 

Fig.~\ref{fig:models} compares surface density profiles obtained for the different stellar distributions in a Hernquist DM halo of $M_h=10^9$ $M_{\odot}$ and scale radius $r_s=1$ kpc. The central stellar density is fixed at $\Sigma_0=3\times10^7M_{\odot}\text{kpc}^{-2}$, and the characteristic radius $R_{1/2}=0.1$ kpc, defined as $\Sigma(R_{1/2})=\Sigma_0/2$. For each distribution we also show the corresponding 3D volumetric density, the predicted line-of-sight velocity dispersion and, finally, the phase-space DF. 

All projected profiles are strongly cored, with differences stemming from the curvature at higher radius and the tail. 
However, as shown in the bottom right panel, only the Einasto with $\alpha=1$, power-potential and S\'ersic profiles lead to consistent, positive phase-space DFs. On the other hand, the Plummer and the Einasto profile with $\alpha=2$, lead to negative values of $f(E)$, as discussed in~\cite{2023ApJ...954..153S}.

\begin{figure}[t]
    \centering
    \includegraphics[width=\linewidth]{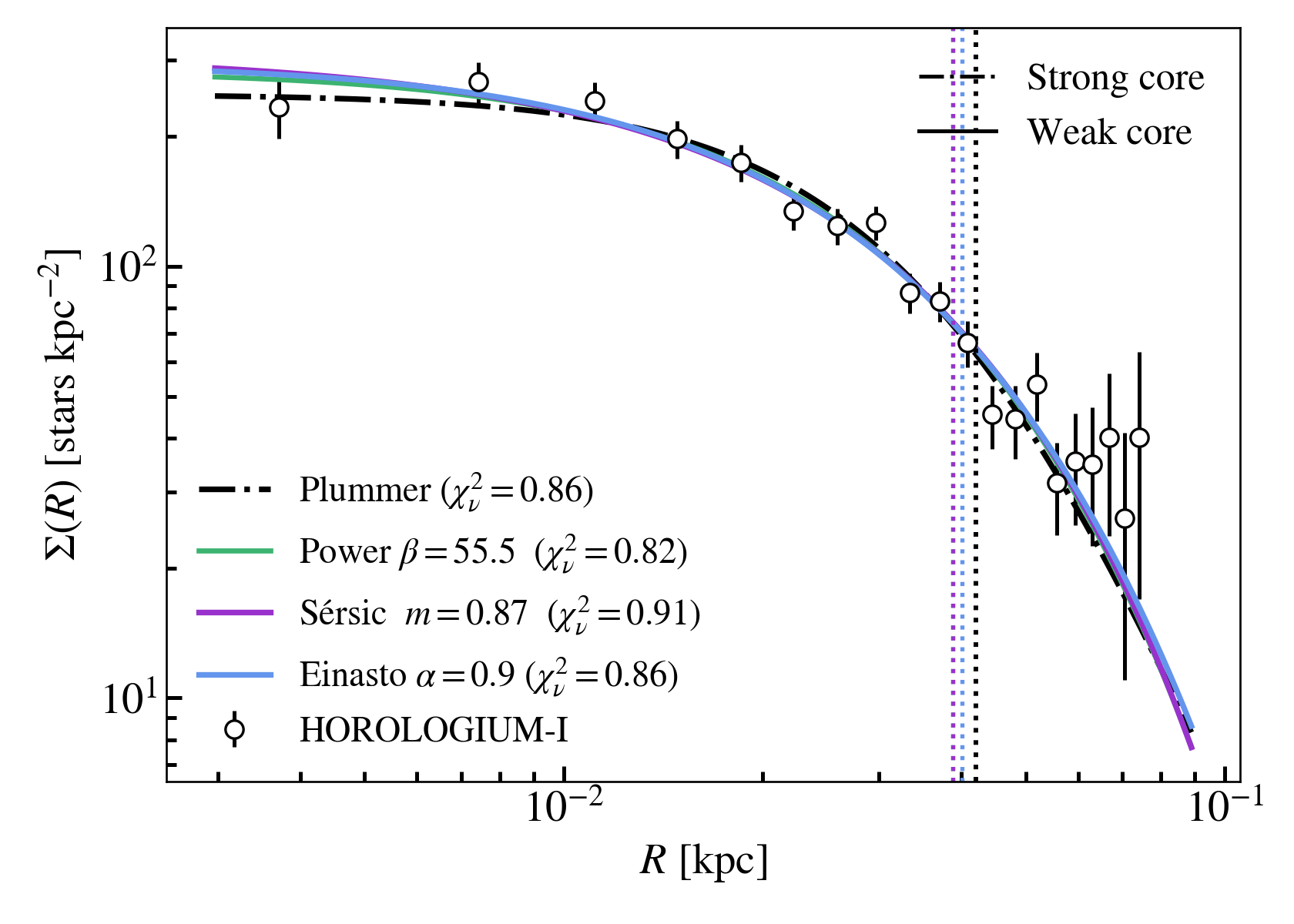}
   \caption{Fits to the surface brightness profile of Horologium I with weak (\textcolor{black}{\rule[0.5ex]{1em}{1pt}}) and strong (\textcolor{black}{%
\rule[0.5ex]{0.4em}{1pt}\hspace{0.15em}%
\raisebox{0.5ex}{.}\hspace{0.15em}%
\rule[0.5ex]{0.4em}{1pt}})
core profiles. See Appendix~\ref{sec:dwarfs_data} for details.
}
    \label{fig:hor-I}
\end{figure}

\subsection{Application to UFDs}
\label{sec:app_UFDs}

Recent observational results from~\cite{2024ApJ...967...72R} have raised the possibility that the stellar cores observed in UFD galaxies may be inconsistent with embeddings within NFW potentials, suggesting a deviation from the collisionless DM paradigm~\citep{2024RNAAS...8..167S}. 
They fit the observed surface density of the UFDs using the shape of the phase-space DF as a free parameter. Here we use a different approach to test this in the context of our classification framework. We re-analyze the stellar surface-density profiles reported in ~\cite{2024ApJ...967...72R}, which studied six UFD satellites of the MW and LMC with deep HST two-band photometry. The dwarfs studied are: Horologium I, Horologium II, Hydra II, Phoenix II, Sagittarius II and Triangulum II, with stellar masses in the range $6\times 10^2-2.4\times 10^4M_{\odot}$. These data are the same as those used in~\cite{2024RNAAS...8..167S} although this reference discussed a stacked surface number density distribution of the six UFDs while in this work we analyze them separately.

Fig. \ref{fig:hor-I} presents the surface number density fits for one representative system, Horologium I, using the models introduced in the previous section (the remaining five fits are provided in Appendix \ref{sec:dwarfs_data}). All models provide a good fit to the data and a weak core is slightly favored over a strong core or weak cusp for the S\'ersic model fit (see Appendix~\ref{sec:dwarfs_data} for the results of the fits and the statistical model comparison). 
The significant uncertainties that typically affect UFDs, due e.g. to the small number of resolved stars, 
then render this preference uncertain.
Despite these limitations, our analysis shows that all examined systems are individually consistent with weakly cored profiles and, thus, with embeddings in cuspy DM halos, as found also by ~\cite{2025arXiv251205719H}.

A key question emerging from this study is whether, given sufficiently deep and statistically robust stellar samples, one can distinguish between weak and strong cores from surface-brightness profile analyses. With this in mind, we turn our attention to dwarf spheroidal (dSph) galaxies of the MW with well-measured surface density profiles.

\subsection{Application to dSph: Fornax}
\label{sec:fornax}

The Fornax dwarf spheroidal, located at a heliocentric distance of $138\pm 8$ kpc \citep{2012A&A...544A..73D}, is one of the best-observed satellites of the MW. It exhibits a prominent stellar core extending well into its central regions \citep{2022MNRAS.512.4171Y}. Dynamical modeling has suggested that Fornax's DM distribution is also cored \citep{2018MNRAS.480..927P, 2019MNRAS.484.1401R, 2020ApJ...904...45H}, a result often interpreted in the context of supernova-driven core formation, expected to be efficient at this stellar-to-halo mass ratio \citep{DiCintio:2013qxa, Tollet:2015gqa}. 
Thus, we analyze the surface brightness profile of Fornax not to directly address the nature of its DM halo, but rather to assess our ability to discriminate between weak and strong stellar cores using its detailed data.

\begin{figure}[t]
    \centering
    \includegraphics[width=0.98\linewidth]{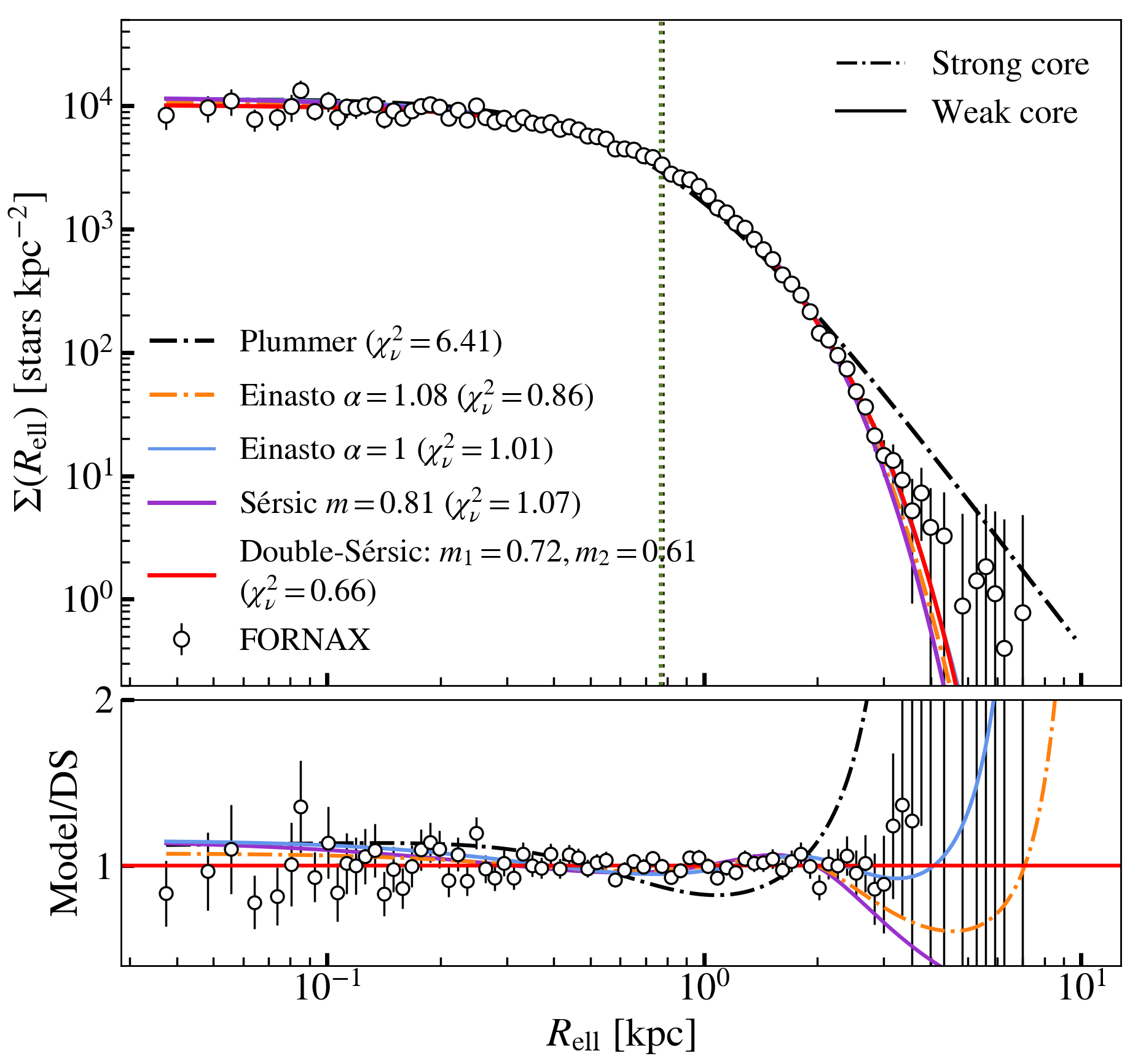}
    \caption{Fornax surface density profile by elliptical radius after background subtraction fitted with different weak     (\textcolor{black}{\rule[0.5ex]{1em}{1pt}}) and strong (\textcolor{black}{%
\rule[0.5ex]{0.4em}{1pt}\hspace{0.15em}%
\raisebox{0.5ex}{.}\hspace{0.15em}%
\rule[0.5ex]{0.4em}{1pt}}) core profiles, index parameter and reduced $\chi^2$ is shown in the plot, the rest of the parameters can be found in Appendix \ref{sec:fornaxfit}.}
    \label{fig:fornax}
\end{figure}

We derive Fornax's surface number density profile from GaiaEDR3 data. Differently from \cite{2022A&A...657A..54B}, we do not calculate probabilities of membership for individual stars, to avoid any sort of bias due to the spatial prior. Rather, we perform a simple selection to remove the obvious contaminants: we retain those sources that have a parallax consistent with zero within 3$\times$ the uncertainty in parallax, proper motion consistent with the systemic proper motion of the galaxy within 3-times the uncertainty (given by square root of the quadratic sum of the proper motion uncertainty for the individual stars and the galaxy's systemic motion) and applying a mask to the (BP-RP) and G colour-magnitude diagram.  Finally, we consider only sources brighter than G=20.5, a regime in which Gaia eDR3 is expected to be 100\% complete at the location of Fornax (estimated using the tools by \cite{2023A&A...669A..55C}). We perform two layers of selections: a ``coarse'' one, with the criteria just described, and a ``strict'' one, where we also apply the quality criteria outlined in Sect.~3 of \cite{2022A&A...657A..54B}. In Fig.~\ref{fig:fornax} we show the fits of several models to Fornax, applied to the ``strict'' selection, but we stress that the results do not change when applied to the other case. 

Moreover, we note that small changes in the binning scheme or in the definition of the background-subtraction region can modify the overall agreement of all models with the data and that we have chosen a scheme with a particularly low $\chi^2_{\nu}$'s. However, the trends and conclusions regarding the ranking of the best-fit models remain unchanged (see Appendix~\ref{sec:fornaxfit} for more details of the fits and discussion of the statistical model comparison).  

A Plummer profile provides a poor fit to the data while a S\'ersic profile provides a good fit with an index $m=0.81$ that corresponds to Fornax having a weak stellar core in 3D. These results are consistent with those obtained in~\cite{2022MNRAS.512.4171Y} as well as with those by ~\citep{2006A&A...459..423B} based on ESO/WFI photometric data. Fitting an Einasto profile one obtains a slightly better fit with an index $\alpha=1.08$ that corresponds to a strong stellar core in 3D, while a better fit is obtained by a sum of S\'ersic profiles (double S\'ersic profile), obtaining results of the two indices which are again consistent with the stellar distribution being a weak core. 

The main conclusion of this analysis is that determining the exact shape of the 3D stellar density profile and the presence of strong cores from number density profiles can be very challenging, even for galaxies with high-quality data such as Fornax. In other words, photometric information alone provides a limited capacity to exclude embeddings of dwarf galaxies in cuspy DM halos. We emphasize, however, that if a non-cored central stellar distribution is observed, this would automatically imply that the stellar 3D density is a cusp.
      
These ambiguities could be alleviated through more robust or model-independent parametrizations of the surface brightness data \citep[see e.g.][]{2025arXiv250926056V,2025ApJ...992..162E}, or by combining them with kinematic constraints, cf. bottom-right panel of Fig.~\ref{fig:models} \citep[see e.g.][]{2018MNRAS.480..927P,2025A&A...699A.347A}. It will be interesting to revisit this issue with future data-sets, as surface number density profiles from missions like Euclid and Nancy Roman.

\section{Conclusions}
\label{sec:conclusions}

We have investigated the connection between the central stellar distributions in dwarf galaxies and the nature of their DM halos, with particular emphasis on the distinction between 3D volumetric densities and their 2D projections when deriving DF consistency conditions via the Eddington inversion. Our main conclusions are:
\begin{enumerate}[label=\roman*., itemsep=0.5em]
    \item 
    We proposed a classification of stellar profiles based on their central density slopes, distinguishing ``weak'' from ``strong'' cores, and ``weak'' from ``strong'' cusps, cf. Tab.~\ref{tab:class} and Fig.~\ref{fig:classification}.
    We showed that \emph{any} spherically symmetric 3D stellar density profile that is a monotonically decreasing function of the radius with a finite central value projects to a \emph{strong core} in 2D. Similarly, 3D weak cusps appear as weak cores in projection, while only strong cusps remain cuspy in 2D. Consequently, \emph{any} 2D surface brightness profile that is not a strong core necessarily corresponds to a cuspy 3D stellar distribution.

    \item
    The sufficient condition for DF negativity based on $d\rho/d\Psi=0$ applies to the \emph{3D density} $\rho(r)$ and potential $\Psi(r)$, not to the projected 2D surface density $\Sigma(R)$. Weak 3D cores embedded in cuspy DM halos can still yield positive $f(E)$ while producing cored surface brightness profiles. Therefore, the presence of a strong central core in projection \emph{does not} rule out a cuspy DM halo.

    \item 
    Using standard stellar distribution models (e.g. Plummer, Einasto, S\'ersic) in Hernquist/NFW DM potentials  and, conversely, building tracers from elemental DFs like $f(E)\propto\delta(E-E_0)$ and $f(E)\propto E^{\beta-3/2}$ (yielding $\rho\propto \Psi^\beta$), we showed explicit families of solutions with strong 2D cores and $f(E)\ge0$ in cuspy DM halos, cf. Fig.~\ref{fig:models}.

    \item Discriminating between cuspy and cored DM profiles using photometric data alone requires going beyond the central flatness of the surface brightness profile. This demands careful analysis of the profile’s curvature through dynamical modeling that enforces DF positivity, properly treats velocity anisotropy, and ideally incorporates kinematic constraints such as velocity dispersions or higher moments. In the case of UFDs analyzed in ~\cite{2024ApJ...967...72R}, with typical stellar masses of $M_\star \sim 10^3-10^4~M_\odot$, fits generally favor weakly cored 3D stellar density profiles over strongly cored ones, cf. Fig.~\ref{fig:hor-I} and Appendix~\ref{sec:dwarfs_data}. We also examined Fornax, which is one of the best-observed MW dSphs, to assess our ability to infer the nature of the 3D stellar distribution from its detailed surface density data. Our analysis shows that recovering the exact shape of the 3D stellar density distribution from cored surface density profiles is highly challenging, cf. Fig.~\ref{fig:fornax}. This suggests that photometric data alone have a limited ability to rule out cuspy DM halos. 

\end{enumerate}

Our analysis is intentionally minimal, assuming spherical symmetry and stationarity. Extending the classification and DF tests to axisymmetric/triaxial systems, non-separable anisotropy, multi-component tracers, and non-equilibrium effects (tides, central heating) will help sharpening the connection between observables and inner DM structure.

\begin{acknowledgements}

We would like to thank Jorge Sánchez Almeida, Ignacio Trujillo, Diego Blas, Kfir Blum, Andrea Caputo and Luca Teodori for reading the manuscript and useful discussions. 
FV, JMC, ADC and GB acknowledge support from the European Union through the grant ``UNDARK'' of the Widening participation and spreading excellence programme (project number 101159929). 
FV and JMC also acknowledge the MICINN through the grant ``DarkMaps'' PID2022-142142NB-I00. 
ADC acknowledges financial support  through the Agencia Estatal de Investigación, 2023 call "Consolidación Investigadora", grant number CNS2023-144669, proyecto "TINY". 
GB acknowledges support from the MCIU/AEI  under grant "FOGALERA" and ERDF with reference PID2023-150319NB-C21/10.13039/501100011033.
RE acknowledges support from the National Science Foundation (NSF) grant AST-2206046. This material is based upon work supported by the National Aeronautics and Space Administration under Grant/Agreement No. 80NSSC24K0084 as part of the Roman Large Wide Field Science program funded through ROSES call NNH22ZDA001N-ROMAN. 
JIR would like to acknowledge support from STFC grants ST/Y002865/1 and ST/Y002857/1. 
\end{acknowledgements}

\bibliographystyle{aa} 

\bibliography{bib}

\appendix

\section{Asymptotic behavior of projected profiles}
\label{app:math_vitamins}

In this section, we derive the asymptotic behavior of the projected surface density profiles around $R\approx0$, presented in Sec.~\ref{sec:cores&cusps}. We start from a volumetric density profile $\rho(r)$, which is spherically symmetric and a monotonically decreasing function of the radius $r$. In addition, we assume that the distribution is integrable and, for definiteness, we impose this condition by introducing a cut-off $r_{\rm max}$ such that $\rho(r)=0$ for $r>r_{\rm max}$. We want to calculate the asymptotic behavior of the Abel projection,
\begin{equation}
\label{eq:projSigma2}\Sigma(R)=2\int^{r_{\rm max}}_Rdr\frac{\rho(r)r}{\sqrt{r^2-R^2}}=2\int_0^{r_{\rm max}}\rho(\sqrt{R^2+l^2})\,dl,
\end{equation}
around $R\approx 0$, and for the different classes of volumetric profiles.

We start projecting the distributions that are finite at the center and can be approximated in its vicinity as in Eq.~\eqref{eq:expansion1} with $\alpha>0$ (i.e. the cores). The first derivative at the center is,
\begin{align}
\label{eq:Sigmaprime}
\Sigma'(0)=\lim_{R\to 0}\;2 R\int_0^{r_{\rm max}} \frac{\rho'(\sqrt{R^2+l^2})}{\sqrt{R^2+l^2}}dl.
\end{align}
and it vanishes unless the integral has a divergence stronger than $\sim 1/R$. The integral is well behaved and bounded for $l>0$ so that such a divergence may only arise in the $l\to 0$ region. We analyze first the case where $\alpha\neq1$, 
\begin{align}
\label{eq:SigmaprimeR0}
\Sigma'(0)&\approx-\lim_{R\to 0}\;2 R\,\alpha a\int_0^{r_{\rm max}}\left(R^2+l^2\right)^{\frac{\alpha}{2}-1}dl\nonumber \\
&\approx-\lim_{R\to 0}\frac{a\alpha\sqrt{\pi}\,\Gamma\left(\frac{1}{2}-\frac{\alpha}{2}\right)}{\Gamma\left(1-\frac{\alpha}{2}\right)}R^\alpha=0,
\end{align}
where we have neglected a $\alpha$-dependent constant resulting from the integral. This constant diverges for the case $\alpha=1$ which requires a separate treatment,
\begin{align}
\label{eq:SigmaprimeR1}
\Sigma'(0)&\approx-\lim_{R\to 0}\;2 R\,a\int_0^{r_{\rm max}}\frac{1}{\sqrt{R^2+l^2}}dl\nonumber \\
&\approx-\lim_{R\to 0}2a\, R\,\log\left(\frac{2r_{\rm max}}{R}\right)=0.
\end{align}
This demonstrates Eqs.~\eqref{eq:bRs_cores} and proves our corollary that all volumetric distributions finite at $r=0$ are strong cores in 2D. 

In the case of cores, integrability implies that the surface density profile is finite at $R\to 0$. However, for cusps boundedness of $\Sigma(0)$ depends on the value of $\gamma$, delimiting the classification in weak and strong cusps,
\begin{align}
\label{eq:SigmaCusp01}
\Sigma(0)&\approx\lim_{R\to0}2\rho_s\int^{r_{\rm max}}_0\left(l^2+R^2\right)^{-\frac{\gamma}{2}}\nonumber\\
&=\lim_{R\to0}\left(C+\frac{\rho_s\sqrt{\pi}\,\Gamma\left(\frac{\gamma}{2}-\frac{1}{2}\right)}{\Gamma\left(\frac{\gamma}{2}\right)}R^{\gamma-1}\right), 
\end{align}
where $C$ is a constant. For strong cusps with $\gamma>1$, $\Sigma(R)$ diverges as $\approx 1/R^{\gamma-1}$ at the center and the projected distribution is a cusp (weak or strong depending on the value of $\gamma$). In case of weak cusps with $0<\gamma<1$, $\Sigma(0)$ is finite and, therefore, the projected distribution is a core in 2D. To prove that it is a weak core, we obtain the slope from Eq.~\eqref{eq:SigmaprimeR0},  so we recover the first two terms in the expansion for this case in Eq.~\eqref{eq:sigmas_sing01}. 

For the limiting case of the weak logarithmic cusp, with $\gamma=0$, we obtain for the expansion around $R\approx0$,
\begin{align}
\label{SigmaCusp0}
\Sigma(R)\approx2\rho_S\int^{r_{\rm max}}_0\log\left(\frac{1}{\sqrt{R^2+l^2}}\right)dl=C-\pi\rho_sR+\mathcal O(R^2),
\end{align}
which is equivalent to Eq.~\eqref{eq:sigmas_sing0}. For the limiting case of the strong cusp with $\gamma=1$, we obtain,
\begin{align}
\label{SigmaCusp1}
\Sigma(0)\approx\lim_{R\to0}2\rho_S\int^{r_{\rm max}}_0 \frac{1}{\sqrt{R^2+l^2}}dl=\lim_{R\to0}\log\left(\frac{2r_{\rm max}}{R}\right),
\end{align}
which diverges logarithmically, in accordance with Eq.~\eqref{eq:sigmas_sing1}, and $\Sigma(R)$ is a weak cusp in 2D.

\section{The positivity of the DF and $d^2\rho/d\Psi^2$}
\label{sec:d2rho}

\begin{figure*}[!hbt]
    \centering
    \includegraphics[width=0.83\linewidth]{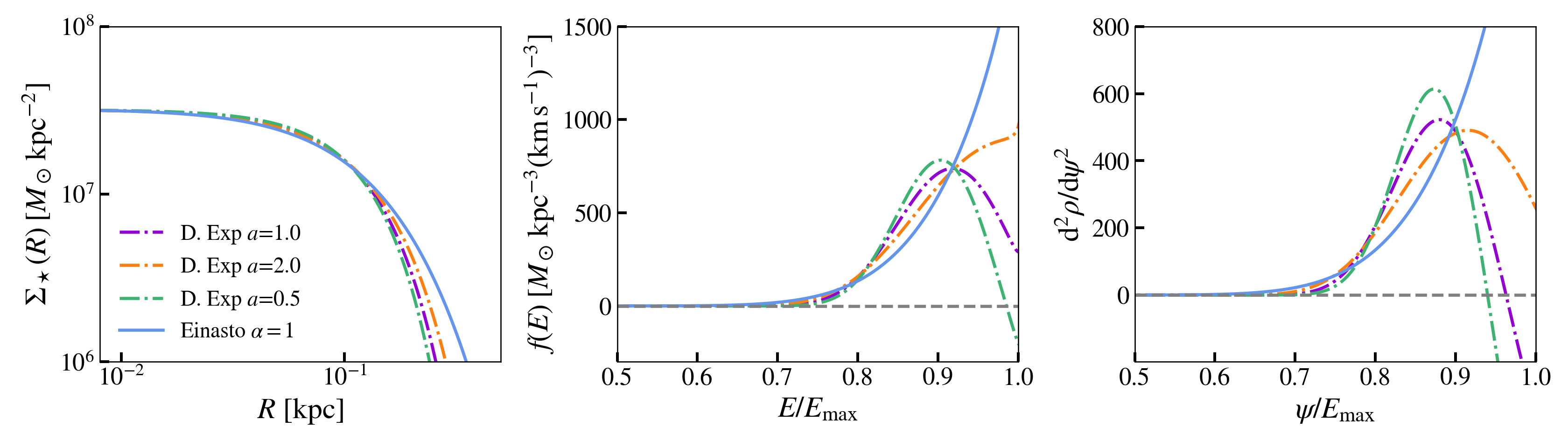}
    \caption{Surface density profile (left), isotropic DF (middle) and $d^2\rho/d\Psi^2$ (right) of a Double-Exponential stellar profile as defined in Eq. \ref{eq:dexp}. Sharp transition parameter between the inner and outer slope result in a marked decrease of $d^2\rho/d\Psi^2$, which may drive the DF to negative values.}
    \label{fig:d2rho_doubleeinasto}
\end{figure*}

As discussed in Sec.~\ref{sec:formalism}, in addition to the central-core condition that distinguishes
between strong and weak cores, the positivity of the distribution function (DF) also depends on the 
\emph{global} shape of the density profile. In particular, the transition from the inner slope to the
outer slope plays a crucial role. Even a profile that exhibits a weak core at small radii may lead to a
negative DF if this transition is sufficiently sharp.

To illustrate this effect, consider a double–Einasto density profile of the form
\begin{equation}
\label{eq:dexp}
  \rho_\star(r) = 
  \rho_0 \exp\!\left[-\left(\frac{r}{r_1}\right)^{\alpha_1}\right]
                  \exp\!\left[-\left(\frac{r}{r_2}\right)^{\alpha_2}\right],
\end{equation}
where $(\alpha_1,r_1)$ describe the inner component and $(\alpha_2,r_2)$ the outer one. Fixing
$\alpha_1 = 1$ and $\alpha_2 = 2$, the behavior near the center corresponds to a weak core.
However, the subsequent transition to the steeper outer component can become arbitrarily sharp,
depending on the ratio of scale radii
\[
    a \equiv \frac{r_1}{r_2}.
\]
For small values of $a$, the outer scale radius $r_2$ becomes large, producing an abrupt change in
slope. In this regime, the quantity $d^2\rho/d\Psi^2$ can become negative, which directly implies a
negative DF. Conversely, for $a>1$ the transition is smoother, and the DF remains positive.
This example highlights that the existence of a weak core is not sufficient to guarantee DF
positivity: the global curvature of the profile, especially the inner-to-outer transition, is equally
determinant.

\section{Extension to anisotropic models}
\label{sec:anisotropy}

When the tracer population is allowed to depart from purely isotropic orbits, the DF becomes a function of both energy and angular momentum, and the orbital structure is conveniently described by the anisotropy parameter
\begin{equation}
    \label{eq:anisotropy}
    \beta=1-\frac{\sigma_{\theta}^2+\sigma_{\phi}^2}{2\sigma_r^2}
\end{equation}
where $\sigma_r$ is the radial velocity dispersion and $\sigma_{\theta}$, $\sigma_{\phi}$ the tangential velocity dispersion in spherical coordinates. Therefore, $\beta=0$ corresponds to isotropy, $\beta>0$ to radially biased orbits, and $\beta<0$ to tangential bias.

An important constraint in anisotropic systems is the central slope-anisotropy theorem~\citep{2006ApJ...642..752A}, which requires for a stellar cusp $\rho_\star \propto r^{-\gamma}$ that $\gamma \geq 2\beta_c$ (with $\beta_c=\lim_{r\to 0}\beta(r)$) for the DF to remain positive. This result generalizes to the global inequality $\gamma(r)\geq 2\beta(r)$ in separable spherical systems~\citep{2010AIPC.1242..300C,2011ApJ...726...80V}. Thus, 3D cores cannot support central radial anisotropy, while steeper cusps allow for stronger radial bias~\citep{2007A&A...471..419B,2009MNRAS.393L..50E}.

A commonly used prescription is the Osipkov-Merritt model~\citep{1979SvAL....5...42O,1985AJ.....90.1027M}, where $\beta(r)$ transitions from isotropy to radial bias given an anisotropy radius $r_a$ that sets the transition. 
This model depends on the angular momentum $L$ and $E=\Phi-v^2/2$, being $\Phi$ the gravitational potential and $v$ the velocity through $Q=E-L^2/(2r_a^2)$, where $r_a$ is a scale radius that defines the transition from an isotropic velocity dispersion at small radii to a radially biased dispersion at large radii. To solve the DF $f(E,L)=f(Q)$ we first integrate $f(Q)$ over the velocity space to obtain the stellar-mass density profile
\begin{equation}
    \rho_\star(r)=\frac{4\pi}{1+r^2/r_a^2}\int^{\phi}_0dQf(Q)\sqrt{2(\phi-Q)}
\end{equation}
The final DF $f(Q)$ can be obtained by Abel transforming $\rho_*$,
\begin{equation}
    f(Q)=\frac{1}{2\pi\sqrt{
    2}}\frac{dG(Q)}{dQ},
\end{equation}
\begin{equation}
    \text{where}\hspace{3mm} G(Q)=-\int^Q_0\frac{d\rho_Q}{d\phi}\frac{d\phi}{\sqrt{Q-\phi}}
\end{equation}
\begin{equation}
    \text{and} \hspace{3mm} \rho(Q)=\left(1+\frac{r^2}{r_a^2}\right)\rho_*(r)
\end{equation}

\begin{figure}[ht!]
    \centering
    \includegraphics[width=\linewidth]{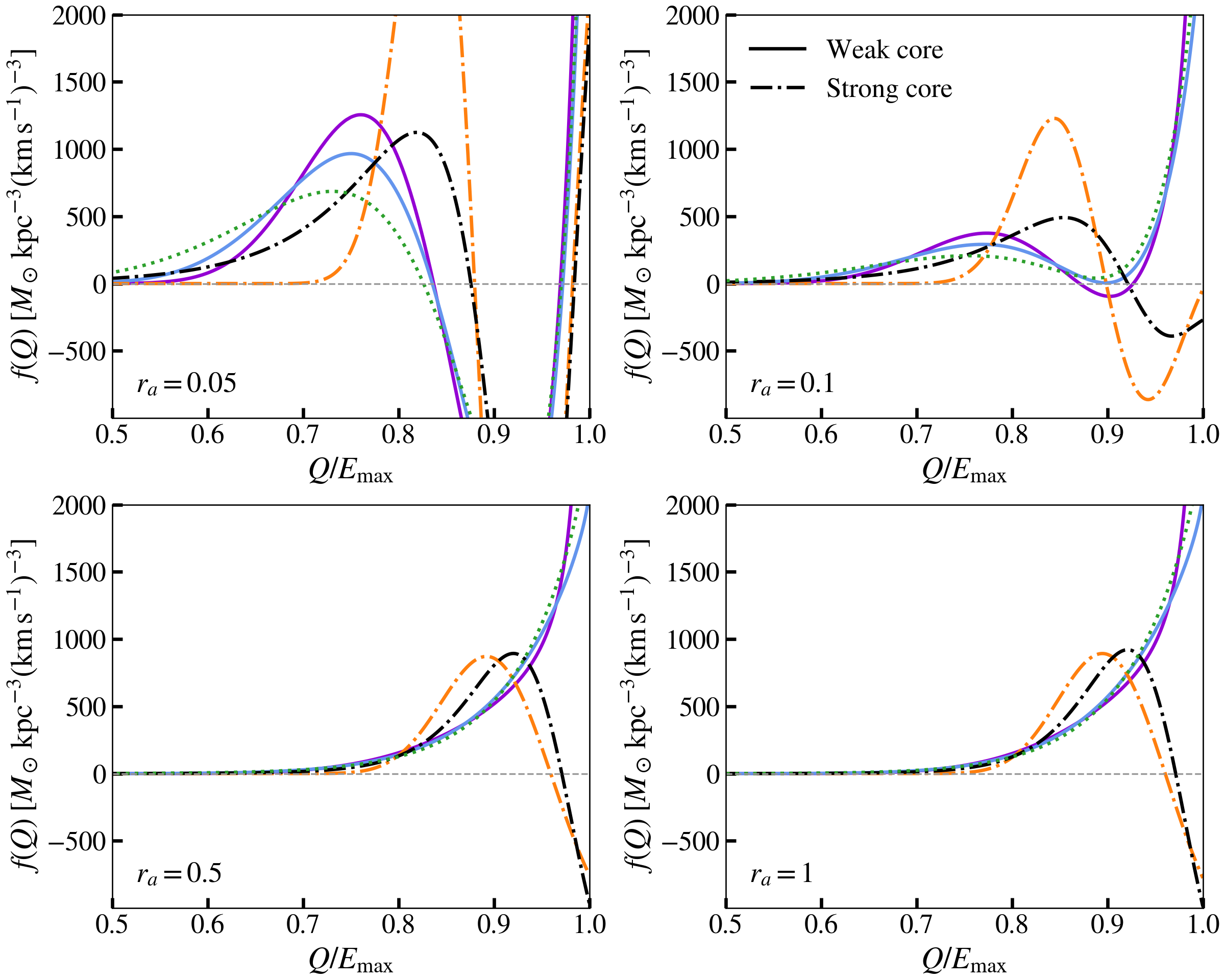}
    \caption{Osipkov--Merritt distribution functions $f(Q)$ for the tracer models considered, shown versus anisotropy radius $r_a$. For strong radial anisotropy ($r_a \ll R_{1/2}$) the DF becomes negative for both weak- and strong-core tracers—consistent with the slope–anisotropy inequality—and is therefore unphysical. Once $r_a \gtrsim R_{1/2}$ (with $R_{1/2}$ the projected half-light radius), $f(Q)$ remains non-negative over the energy range, making the models compatible with the OM parameterization.}
    \label{fig:OMfQ}
\end{figure}

\section{NFW}
\label{app:NFW}
\begin{figure}[hb!]
    \centering
    \includegraphics[width=\linewidth]{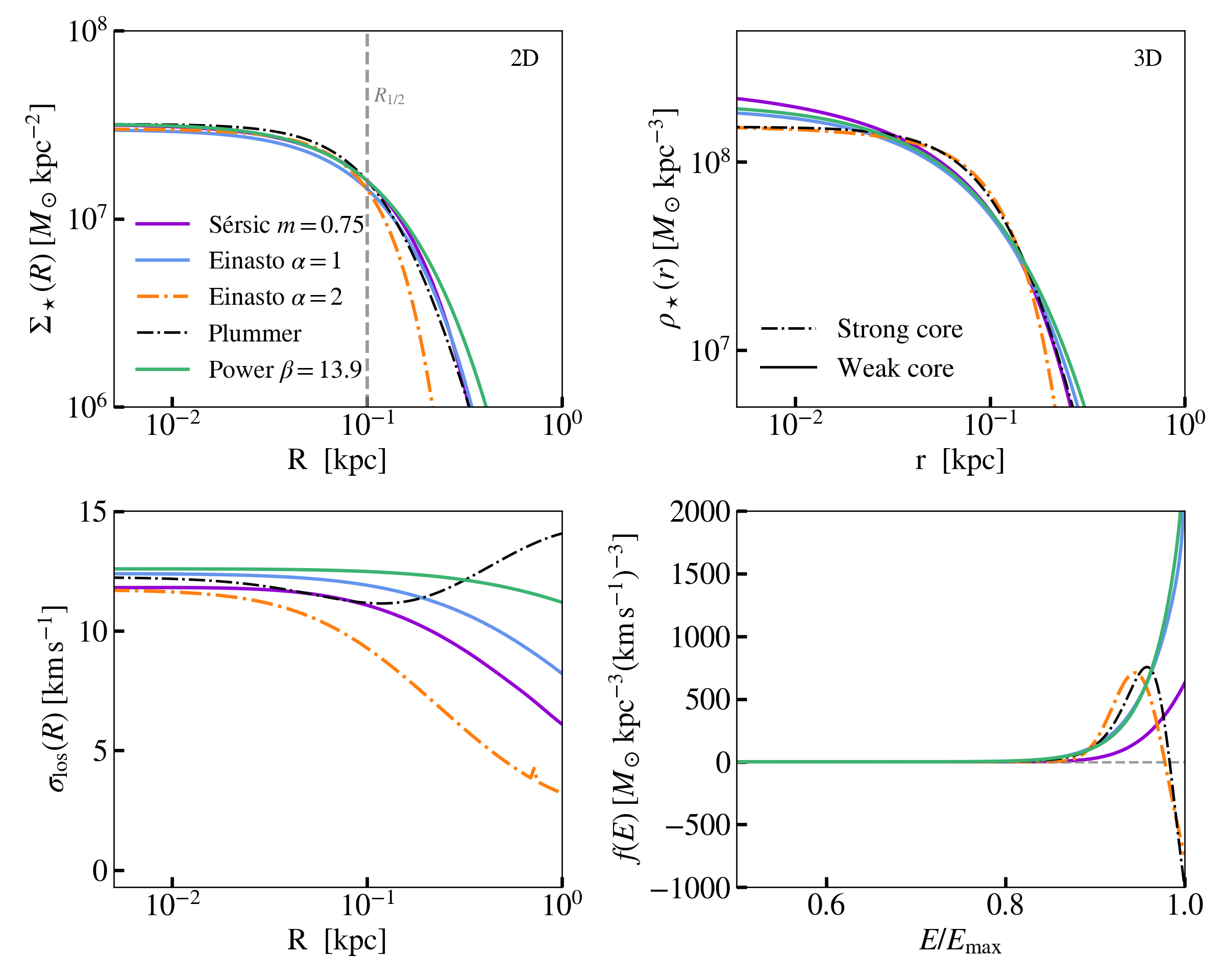}
    \caption{Same as Fig. \ref{fig:models} with an NFW DM potential of  $M_{200}=10^9M_{\odot}$, $r_s=1$kpc. Stellar tracers: $\Sigma_0=3\times10^7M_{\odot}\text{kpc}^{-2}$, $R_{1/2}=0.1$kpc. }
    \label{fig:DFiNNFW}
\end{figure}

The conclusions above do not depend on the cuspy profile we choose. Hernquist is a convenient choice to derive the results analytically from the bottom-up approach of the power-potential DF. While the canonical NFW potential 
\begin{equation}
    \rho_{\rm NFW}(r)=\frac{\rho_s}{(r/r_s)(1+r/r_s)^2}
\end{equation}
introduces a logarithmically divergent potential term at small radii, treating it numerically leaves our conclusions unchanged. Fig. \ref{fig:DFiNNFW} shows the same quantities as Fig. \ref{fig:models} but the tracer embedded in a cosmological NFW with $M_{200}=10^9 M_{\odot}$ and following a $c-M$ relation.

\clearpage
\onecolumn

\section{Individual fits to UFD}
\label{sec:dwarfs_data}

\begin{figure*}[h]
    \centering
    \includegraphics[width=\linewidth]{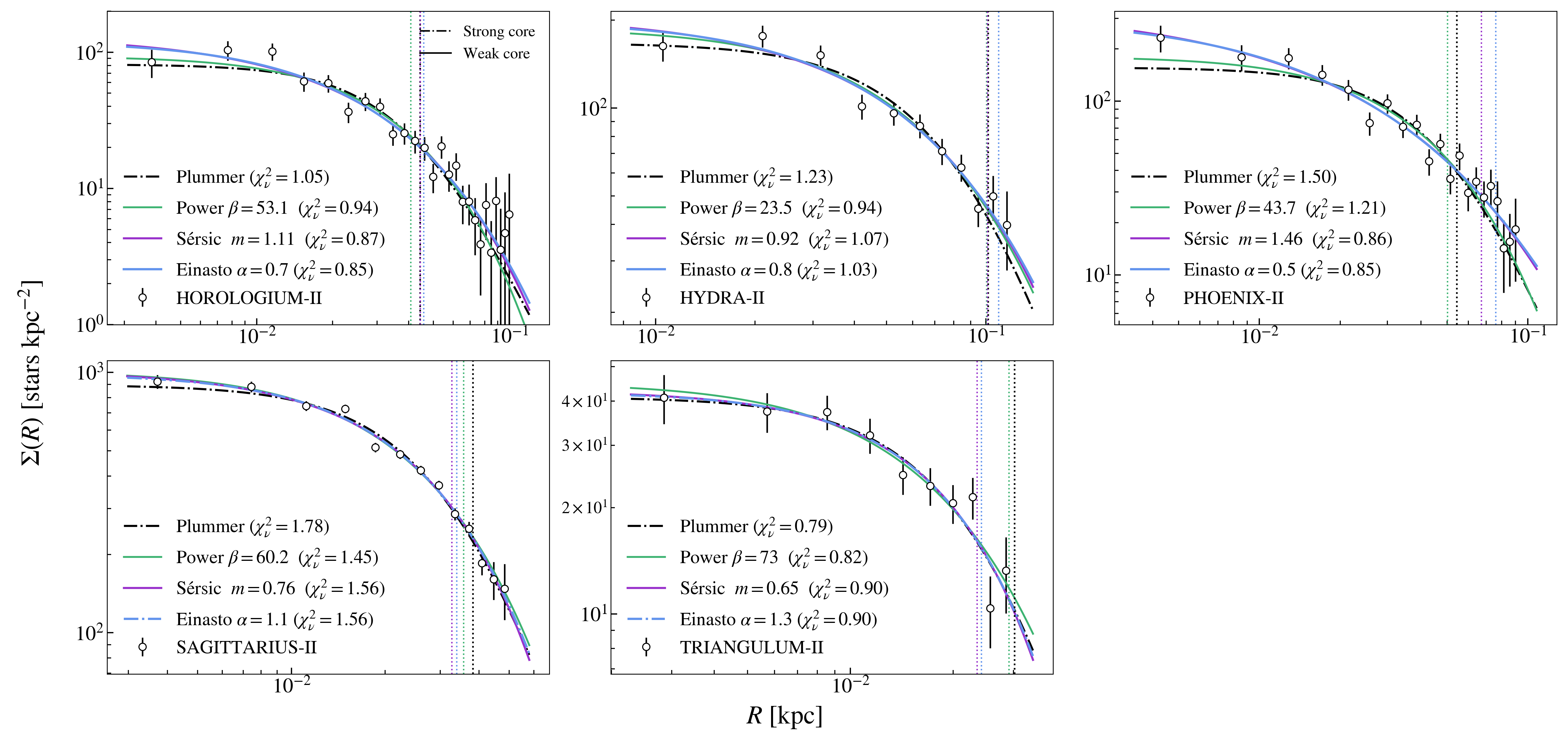}
    \caption{Fits to the surface number density profile of 5 UFDs using the Plummer, Einasto, and power-potential models. In the Plummer model, both the normalization and the scale radius are treated as free parameters. For the Einasto model, we additionally allow the shape parameter $\alpha$ to vary. In the power-potential model, the normalization and the slope parameter $\beta$ are free, while $r_s$ is fixed to the dark-matter halo scale, assumed here to be 1 kpc.}
    \label{fig:ufds}
\end{figure*}

For the fits to UFDs, in the Plummer model, both the normalization and the scale radius are treated as free parameters. For the Einasto model, we additionally allow the shape parameter $\alpha$ to vary between strong core ($\alpha>1$) or weak core ($0<\alpha\leq1$). In the power-potential model, the normalization and the slope parameter $\beta$ are free, while $r_s$ is fixed to the dark-matter halo scale, assumed here to be 1 kpc. Fig. \ref{fig:ufds} and Table \ref{tab:ufds}
 summarizes the data and fits performed to the UFD data.

\begin{table*}[h]
\centering
\small
\caption{Best-fit structural parameters for the surface number density profiles of the UFD sample.}
\label{tab:ufds}
\begin{tabular}{lcccccc}
\hline\hline
\rule{0pt}{1em}
Model / Parameter & Hor~I & Hor~II & Hydra~II & Phoe~II & Sgr~II & Tri~II \\[0.3em]
\hline
\rule{0pt}{1.1em}
\textbf{Plummer} & & & & & & \\
$A$ [stars kpc$^{-2}$] & 251$\pm$14 & 81.4$\pm$6.8 & 167$\pm$9 & 156$\pm$12 & 895$\pm$24 & 41$\pm$3.1 \\
$a$ [kpc] & 0.042$\pm$0.002 & 0.0444$\pm$0.0024 & 0.102$\pm$0.006 & 0.0545$\pm$0.0031 & 0.0382$\pm$0.0009 & 0.0302$\pm$0.0025 \\
$\chi^2_\nu$ & 0.86 & 1.05 & 1.23 & 1.50 & 1.78 & 0.79 \\
AIC$_c$ & 20.20 & 29.74 & 16.58 & 33.01 & 24.76 & 12.068\\[0.6em]

\textbf{Einasto} & & & & & & \\
$A$ [stars kpc$^{-2}$] & 295$\pm$33 & 120$\pm$23 & 199$\pm$27 & 298$\pm$75 & 987$\pm$52 & 42.3$\pm$5.6 \\
$r_{-2}$ [kpc] & 0.0376$\pm$0.0018 & 0.0373$\pm$0.0026 & 0.0974$\pm$0.0116 & 0.0478$\pm$0.0045 & 0.0341$\pm$0.0011 & 0.0262$\pm$0.0035 \\
$\alpha$ & 0.890$\pm$0.18 & 0.675$\pm$0.15 & 0.797$\pm$0.28 & 0.463$\pm$0.13 & 1.051$\pm$0.13 & 1.318$\pm$0.45 \\
$\chi^2_\nu$ & 0.86 & 0.86 & 1.03 & 0.85 & 1.56 & 0.90 \\
AIC$_c$ & 22.17 & 26.75 & 17.68 & 22.74 & 24.27 & 16.29\\[0.6em]

\textbf{Power-potential} & & & & & & \\
$A$ [stars kpc$^{-2}$] & 284$\pm$16 & 93$\pm$8 & 188$\pm$12 & 180$\pm$14 & 1010$\pm$29 & 45$\pm$4 \\
$\beta$ & 55.5$\pm$2.5 & 53.1$\pm$2.7 & 23.5$\pm$1.4 & 43.7$\pm$2.5 & 60.2$\pm$1.6 & 73$\pm$7 \\
$\chi^2_\nu$ & 0.82 & 0.94 & 0.94 & 1.21 & 1.45 & 0.82 \\
AIC$_c$ & 19.46 & 27.06 & 13.96 & 27.57 & 21.24 & 12.29\\[0.6em]

\textbf{Sérsic} & & & & & & \\
$I_0$ [stars kpc$^{-2}$] & 311$\pm$41 & 133$\pm$30 & 208$\pm$34 & 351$\pm$114 & 1020$\pm$63 & 42.9$\pm$6.5 \\
$R_s$ [kpc] & 0.0284$\pm$0.0038 & 0.022$\pm$0.006 & 0.0693$\pm$0.0118 & 0.0175$\pm$0.0084 & 0.0282$\pm$0.0015 & 0.0238$\pm$0.0026 \\
$m$ & 0.875$\pm$0.132 & 1.11$\pm$0.19 & 0.92$\pm$0.23 & 1.46$\pm$0.34 & 0.77$\pm$0.07 & 0.65$\pm$0.22 \\
$\chi^2_\nu$ & 0.91 & 0.87 & 1.07 & 0.86 & 1.56 & 0.90 \\
AIC$_c$ & 22.95 & 27.09 & 17.97 & 22.83 & 24.26 & 16.28\\[0.6em]
\hline
\end{tabular}
\end{table*}

\twocolumn
In Table~\ref{tab:ufds} we report the corrected Akaike Information Criterion (AIC$_c$) as a quantitative metric for model comparison. The standard definition is \citep{2007MNRAS.377L..74L}
\begin{equation}
    \mathrm{AIC}_c = -2\ln\mathcal{L}_{\max} + 2k + \frac{2k(k+1)}{N-k-1},
\end{equation}
where $\mathcal{L}_{\max}$ is the maximum likelihood, $k$ the number of free parameters, and $N$ the number of data points.

Since our fits assume Gaussian errors, the likelihood and chi--square are related by
\begin{equation}
    -2\ln\mathcal{L}_{\max} = \chi^2 + \text{constant},
\end{equation}
and the additive constant cancels when comparing models. Therefore, in our analysis we use the equivalent expression
\begin{equation}
    \mathrm{AIC}_c = \chi^2 + 2k + \frac{2k(k+1)}{N-k-1}.
\end{equation}
The preferred model is the one that yields the lowest AIC$_c$.

\section{Fit to Fornax}
\label{sec:fornaxfit}
\begin{figure}[hb!]
    \centering
    \includegraphics[width=\linewidth]{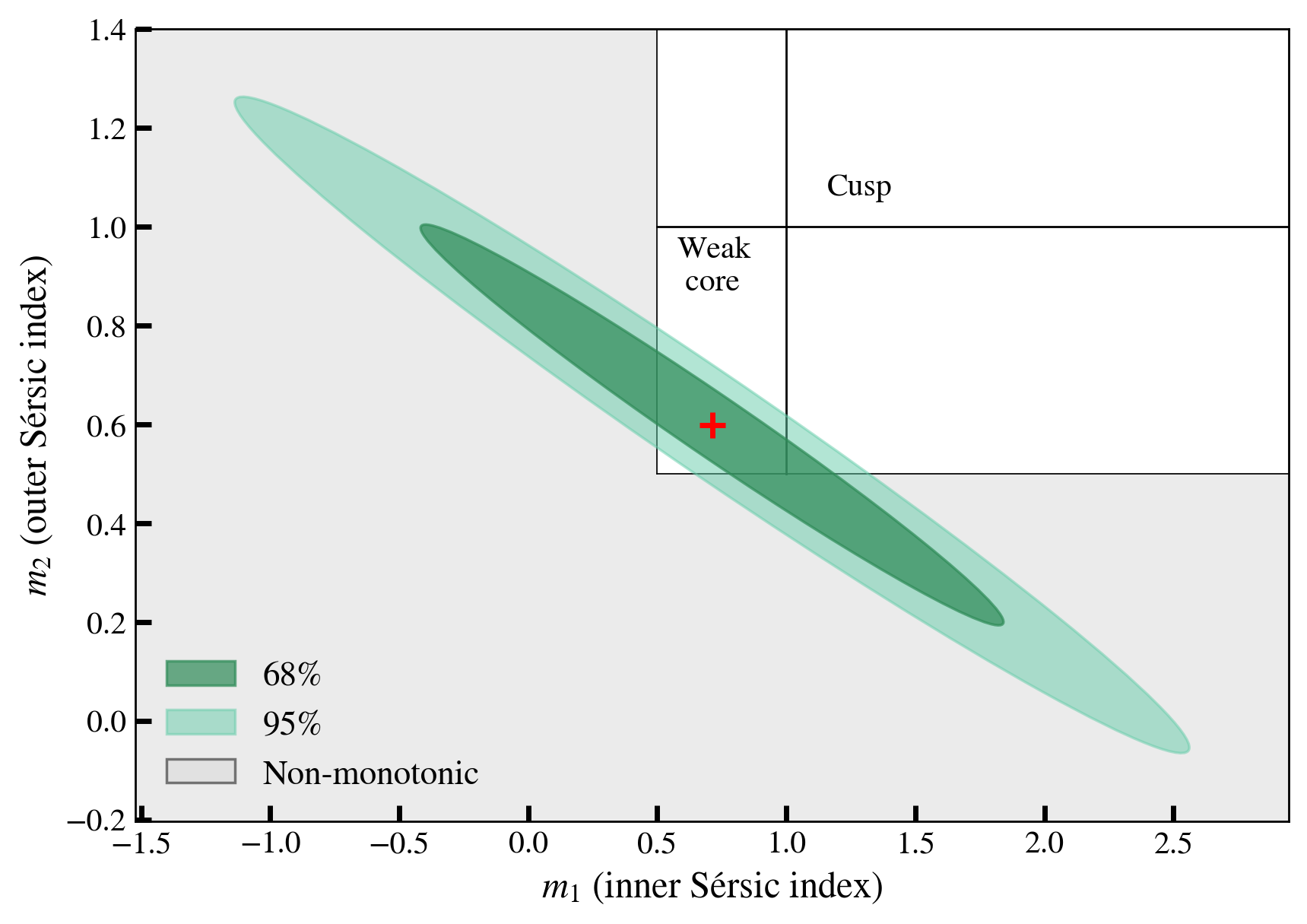}
    \caption{Corner plot illustrating the joint posterior distributions of the parameters from the double-Sérsic fit to Fornax. The best-fit parameters lie within the weak-core regime of the accessible Sérsic indices.}
    \label{fig:corner}
\end{figure}

Table \ref{tab:fornax} lists the best-fit parameters and their corresponding uncertainties obtained from the fits to Fornax described in Sec. \ref{sec:fornax}, while Fig. \ref{fig:corner} shows a corner plot of the best-fit Sérsic indices for the double-Sérsic model.

\begin{table}[h!]
\centering
\small
\caption{Best-fit structural parameters for Fornax and the different surface-brightness density models.}
\label{tab:fornax}
\begin{tabular}{lccccc}
\hline
\hline
\rule{0pt}{1em}
Model & \multicolumn{3}{c}{Parameters} & $\chi^2/\nu$ & AICc \\[0.4em]
\hline
\rule{0pt}{1.1em}

Plummer & $\Sigma_0$ & $a$ & & \\
& \footnotesize{[$10^4$ stars kpc$^{-2}$] }& [kpc] & & \\
& 1.14$\pm$0.02 & 0.78$\pm$0.01 & & 6.41 & 510.63 \\[1em]

Sérsic & $I_0$ & $R_s$ & $m$ & \\
& \footnotesize{[$10^4$ stars kpc$^{-2}$] }& [kpc] & & \\
& 1.18$\pm$0.03 & 0.63$\pm$0.01 & 0.81$\pm$0.01 & 1.07 & 89.77 \\[1em]

Einasto & $\rho_0$ & $r_{-2}$ & $\alpha$ \\
& \footnotesize{[$10^3$ stars kpc$^{-3}$] }& [kpc] & \\
& 2.00$\pm$0.05 & 0.78$\pm$0.01 & 1.08$\pm$0.02 & 0.87 & 73.77 \\[1em]

Einasto & $\rho_0$ & $r_{-2}$ & \\
($\alpha=1) $& \footnotesize{[$10^3$ stars kpc$^{-3}$] }& [kpc] &  \\
& 2.10$\pm$0.04 & 0.76$\pm$0.00 & & 1.01 & 83.99 \\[1em]

Double-  & $I_{0,i}$ & $R_{s,i}$ & $m_i$ & \\
Sérsic& \footnotesize{[$10^3$ stars kpc$^{-2}$] }& [kpc] & & \\
\textit{(inner)} & 3.21$\pm$18.13 & 0.90$\pm$1.89 & 0.72$\pm$0.80 & \multirow{2}{*}{0.66} & \multirow{2}{*}{62.77} \\[0.2em]
\textit{(outer)} & 6.97$\pm$17.50 & 0.66$\pm$0.11 & 0.61$\pm$0.27 & & \\[0.3em]

\hline
\end{tabular}
\end{table}

\end{document}